www.freewebs.com/vpsawana

# MAGNETOTRANSPORT OF $La_{0.70}Ca_{0.3-x}Sr_xMnO_3$ (Ag): A POTENTIAL ROOM TEMPERATURE BOLOMETER AND MAGNETIC SENSOR

V.P.S. Awana[1,*], Rahul Tripathi, Neeraj Kumar, and H. Kishan

National Physical Laboratory, Dr. K.S. Krishnan Marg, New Delhi 110012, India

G.L. Bhalla

Department of Physics and Astrophysics, University of Delhi, Delhi-110007, India

R. Zeng

ISEM, University of Wollongong, NSW 2522, Australia

L.S. Sharth Chandra,

UGC-DAE Consortium for Scientific Research, University Campus, Khandwa Road, Indore-452017, India and Material and Advanced Accelerator Science Division, RRCAT, Indore-452013, India

V. Ganesan

UGC-DAE Consortium for Scientific Research, University Campus, Khandwa Road, Indore- 452017, India

H.U. Habermeier

Max Planck Institute for Solid State Research, Heisenbergstrasse-1, D-70569, Stuttgart Germany

*: Corresponding Author
Dr. V.P.S. Awana
National Physical Laboratory, Dr. K.S. Krishnan Marg, New Delhi-110012, India
Fax No. 0091-11-45609310: Phone No. 0091-11-45608329
e-mail-awana@mail.nplindia.ernet.in: www.freewebs.com/vpsawana/

## Abstract

Here we report the optimized magneto-transport properties of polycrystalline $La_{0.70}Ca_{0.3-x}Sr_xMnO_3$ and their composites with Ag. The optimization was carried out by varying the Sr and Ag contents simultaneously to achieve large temperature coefficient of resistance (TCR) as well as low field magneto-resistance (MR) at room temperature. Sharpest paramagnetic (PM)-ferromagnetic (FM) and insulator-metal (IM) transition is observed in the vicinity of the room temperature ($T_C \approx 300$ K $\approx T_{IM}$) for the composition $La_{0.70}Ca_{0.20}Sr0_{0.10}MnO_3:Ag_{0.20}$. Partial substitution of larger $Sr^{2+}$ ions at the $Ca^{2+}$ ions sites controls the magnitude of the FM and IM transition temperatures, while the Ag induces the desired sharpness in these transitions. For the optimized composition, maximum TCR and MR are tuned to room temperature (300 K) with the former being as high as 9% and the later being ~20 % and ~30 % at 5 and 10 kOe magnetic fields respectively. Such sharp single peak (TCR~9 %) at room temperature can be used for the bolometric and infrared detector applications. The achievement of large TCR and low field MR at T~300K in polycrystalline samples is encouraging and we believe that further improvements can be achieved in thin films, which, by virtue of their low conduction noise, are more suitable for device applications.

## 1. Introduction

Manganites with general formula $A_{1-x}Á_xMnO_3$ (where A = rare earth element and Á = Alkaline earth metals) are a focal point of research since a long time [1]. These compounds are considered to be promising material for the technological opinion [1-3]. Among all fascinating properties, the most outstanding and extensively explored assets are the magneto-resistance (MR) and temperature coefficient of resistance (TCR). Magneto-resistance (MR) is the relative change in the electrical resistivity by application of magnetic field and similarly TCR is parametric notation of transition sharpness, defined as, $[1/R * (dR / dT)]$, where R, T is the resistance and temperature respectively. It is seen that the maximum MR as well as TCR in hole-doped manganites occur near the metal-insulator (MI) transition ($T_{MI}$) being accompanied with ferromagnetic-paramagnetic (FM) transition. The steep transition about metal-insulator crossover determines the sensitivity as well as active zone for these sensors. Practically one desires to have higher TCR and MR near room temperature, i.e. at 300K.

In this direction there had been various trials before and TCR as high as above 10% and reasonable MR is seen in $La_{2/3}Ca_{1/3}MnO_3$:$Ag_y$ composites [4-6]. However the high TCR and optimized reasonable MR, in particular at low fields is always seen only at below room temperature (< 265K) only [4-7]. As mentioned before for practical use of manganites as sensors, one aspires for high TCR and MR at room temperature. Though, the working temperature could be enhanced to near room temperature in case of $La_{0.7}Ca_{0.3-x}Ba_xMnO_3$, the observed values of TCR and MR are reasonable good, but could not improved a lot with silver addition [8]. In fact our earlier efforts in case of $La_{0.7}Sr_{0.3}MnO_3/La_{0.7}Ba_{0.3}MnO_3$, we were not successful [9]. In view of our consistent approach to improve room temperature TCR and low field magneto resistance (LFMR) in case of pristine [5] and Ba doped [8] silver composites, in present paper we deliberate on Sr doped $La_{0.7}Ca_{0.3}MnO_3$ and their Ag added compounds. For optimized composition of $La_{0.7}Ca_{0.20}Sr_{0.10}MnO_3$:$Ag_{0.2}$, the maximum TCR and MR are tuned to 300 K with the former being as high as 9%/K and the later being above 20% at 5 kOe and 35% at 10 kOe field. Although the studied samples are bulk in nature, the thin films fabricated from these bulk targets will serve the ultimate purpose of manganite being used as potential magnetic sensors at room temperature. Further above 9%/K single peak sharp TCR at 300K can be used for the bolometric and infra-red applications. Improved TCR (>9%) and MR (> 20% at 5 kOe) at 300 K is the maximum yet reported for any manganite at room temperature or above.

## 2. Experimental Details

The samples of the series $La_{0.7}Ca_{0.3-x}Sr_xMnO3$:$Ag_y$ (x = 0.0, 0.05, 010; y = 0, 0.1, 0.2, 0.3 and 0.4 ) are synthesized by solid-state reaction route using ingredients $La_2O_3$, $CaCO_3$, $SrCO_3$, $MnO_2$ and Ag (Powder). The mixed powders were calcined at 1000°C, 1100°C and 1200°C in air for 24 hours and followed by thorough grounding each time.

Then the powders were pre-sintered at 1300°C in air for 24 hours. Finally the pelletized ceramics were annealed in air for 24 hour at 1400°C. For loading of oxygen, the pellets were annealed in the flow of oxygen at 1100°C for 12 hours and subsequently slow cooled to room temperature. The structure and phase purity of the samples were analyzed by powder X–ray diffraction (XRD) taken on Rigaku mini-flex diffrractometer. The R(T) measurements with and without magnetic field (< 14 Tesla) were carried out using four-probe method in the temperature range of 5 to 400K on a Quantum Design PPMS (Physical Property Measurement System).

**3. Results & Discussion**

Fig.1 depicts the room temperature X-ray diffraction (XRD) patterns of $La_{0.7}Ca_{0.25}Sr_{0.05}MnO_3$ (LCS05MO), $La_{0.7}Ca_{0.2}Sr_{0.1}MnO_3$ (LCS10MO), $La_{0.7}Ca_{0.2}Sr_{0.1}MnO_3:Ag_{0.4}$ (LCS10MOAg(0.2)) along with earlier studied $La_{2/3}Ca_{1/3}MnO_3:Ag_{0.4}$ (LCMOAg(0.4) [5]. All the studied samples are crystallized in near single phase along with some silver lines in Ag added compounds. Details of structural refinements along with lattice parameters variations will be reported else where in full paper [10].

The normalized resistance ($R_T/R_{400K}$) plots for various Ca/Sr substituted and Ag added $LaMnO_3$ samples are shown in Fig. 2. As it is well known that $La_{0.7}Sr_{0.3}MnO_3$ has high metal-insulator transition temperature ($T_{MI}$) than $La_{0.7}Ca_{0.3}MnO_3$. This change in the transition temperature with divalent ions (Ca, Sr) substitution is related to the variation in Mn-O-Mn bond angle governs by the ionic size of the dopant. Tilting of Mn-O bond directly affects the electron hopping amplitude from one Mn site to another Mn site in these compounds. So, by proper optimization of Sr and Ca concentration one can easily tune the $T_{MI}$ to room temperature. As shown is Fig.2, the $T_{MI}$ increases with Sr concentration in $La_{0.7}Ca_{0.3-x}Sr_xMnO_3$. But the sharpness of transition decreases that could be recovered by reducing the inter-grain tunnel resistance. As in ferromagnetic/metallic region the inter-grain barrier in granular perovskite behaves as non-magnetic as well as non-conducting. So the double-exchange interaction, which is often used to explain the conductive behavior of manganite, becomes weaker on the surfaces of grains. However with the help of external field or by producing a conducting channel between the grains the tunnel resistance can be reduced. It is believed that if it may possible than granular perovskite can have large value of TCR and can behave like tunnel- GMR. In order to enhance the sharpness/ inter-grain conductivity, silver has been added to the pristine samples. Metallic Silver having low melting point will help during sintering process (Liquid phase sintering) and also segregates at the grain boundaries (GB's) creating conducting channels The SEM micrograph of pure and Ag-doped sample of LCS10MO are shown on Fig.3(a) & 3(b). These figures clearly indicate that sample with silver as additives have better grain growth and connectivity. The $T_{MI}$ remains nearly constant with silver addition [5, 8], though sharpness improves. Inset is showing the TCR (measure of sharpness) as a function of temperature for LCMOAg(0.4) and LCS10MOAg(0.2) composition. The maximum TCR (~15%/K) is obtained for LCMOAg(0.4) but below room temperature. For un-cooled operations, room temperature high TCR is required. A ~9%/K value of TCR above room temperature is obtained for LCS10MOAg (02), which is the optimum value of TCR achieved in the case of $La_{0.7}Ca_{0.2}Sr_{0.1}MnO_3:Ag_y$ compounds. In polycrystalline bulk samples, a ~9%/K value above room temperature is quite reasonable. And further optimization of samples as a thin film can yield good and encouraging results for application purposes.

As far as magnetic measurements are concerned, in $La_{2/3}Ca_{1/3}MnO_3$:Ag system, the para-ferromagnetic transition temperature($T_C$) remains almost constant for all compositions [5]. But $T_C$ varies slightly for $L_{0.7}Ca_{0.2}Sr_{0.1}MnO_3+Ag_y$ (y = 0, 0.1, 0.2, 0.3, 0.4) compounds. The variation in $T_C$ can be correlated with slight substitution of Ag into the main matrix (Sr being larger) or to the concept of mixed phase in manganites [11]. As shown in Fig.4, there is large change in the resistivity values of LCS10MO(Ag0.2) with applied field. The change in the resistivity with applied field has been calculated in terms of MR (inset Fig.3). For application point of view, low field magneto-resistance (LFMR) is required, since high magnetic fields are not easily accessible. In our samples, high MR at low field has been observed for LCS10MO silver samples. For LCS10MOAg(0.2) samples, the observed MR is about ~30% at room temperature under applied field of 10 kOe, which is quite high at room temperature under low field. The maximum MR value for pristine $La_{0.7}Ca_{0.2}Sr_{0.1}MnO_3$ is nearly 16% (10 kOe) and that of LCMOAg(0.4) is 10% (10 kOe) at 300 K. With such a high LFMR values for $La_{0.7}Ca_{0.2}Sr_{0.1}MnO_3$:$Ag_y$, with optimum value at Ag(0.2),can be explored and further optimized for industrial purposes.

In summery we have optimized both high TCR (~ 9 % / K) and LFMR (~ 30% at 10 kOe) at room temperature for $La_{0.7}Ca_{0.2}Sr_{0.1}MnO_3$ + Ag (02) compound. The studied materials could be the potential candidates for room temperature bolometer (high TCR) and magnetic sensor (high MR).

## Figure Captions

Figure 1: XRD Pattern of $La_{2/3}Ca_{1/3}MnO_3$:$Ag_{0.4}$, $La_{0.7}Ca_{0.25}Sr_{0.05}MnO_3$, $La_{0.7}Ca_{0.2}Sr_{0.1}MnO_3$, $La_{0.7}Ca_{0.2}Sr_{0.1}MnO_3$:$Ag_{0.2}$.

Figure 2: Normalized R(T) plots of $La_{2/3}Ca_{1/3}MnO_3$:$Ag_{0.4}$, $La_{0.7}Ca_{0.25}Sr_{0.05}MnO_3$, $La_{0.7}Ca_{0.2}Sr_{0.1}MnO_3$ and $La_{0.7}Ca_{0.2}Sr_{0.1}MnO_3$:$Ag_{0.2}$. Inset shows the TCR% as a function of temperature for $La_{2/3}Ca_{1/3}MnO_3$:$Ag_{0.4}$ and $La_{0.7}Ca_{0.2}Sr_{0.1}MnO_3$:$Ag_{0.2}$.

Figure 3: SEM micrograph of (a) $La_{0.7}Ca_{0.2}Sr_{0.1}MnO_3$ and (b) $La_{0.7}Ca_{0.2}Sr_{0.1}MnO_3$:$Ag_{0.2}$

Figure 4: Resistivity plots of $La_{0.7}Ca_{0.2}Sr_{0.1}MnO_3$:$Ag_{0.2}$ under applied field of 0, 7 T & 14 T. Inset shows MR% of $La_{0.7}Ca_{0.2}Sr_{0.1}MnO_3$:$Ag_{0.2}$ at fixed temperature of 100 K, 200 K, 300 K & 350 K

**Figure 1**

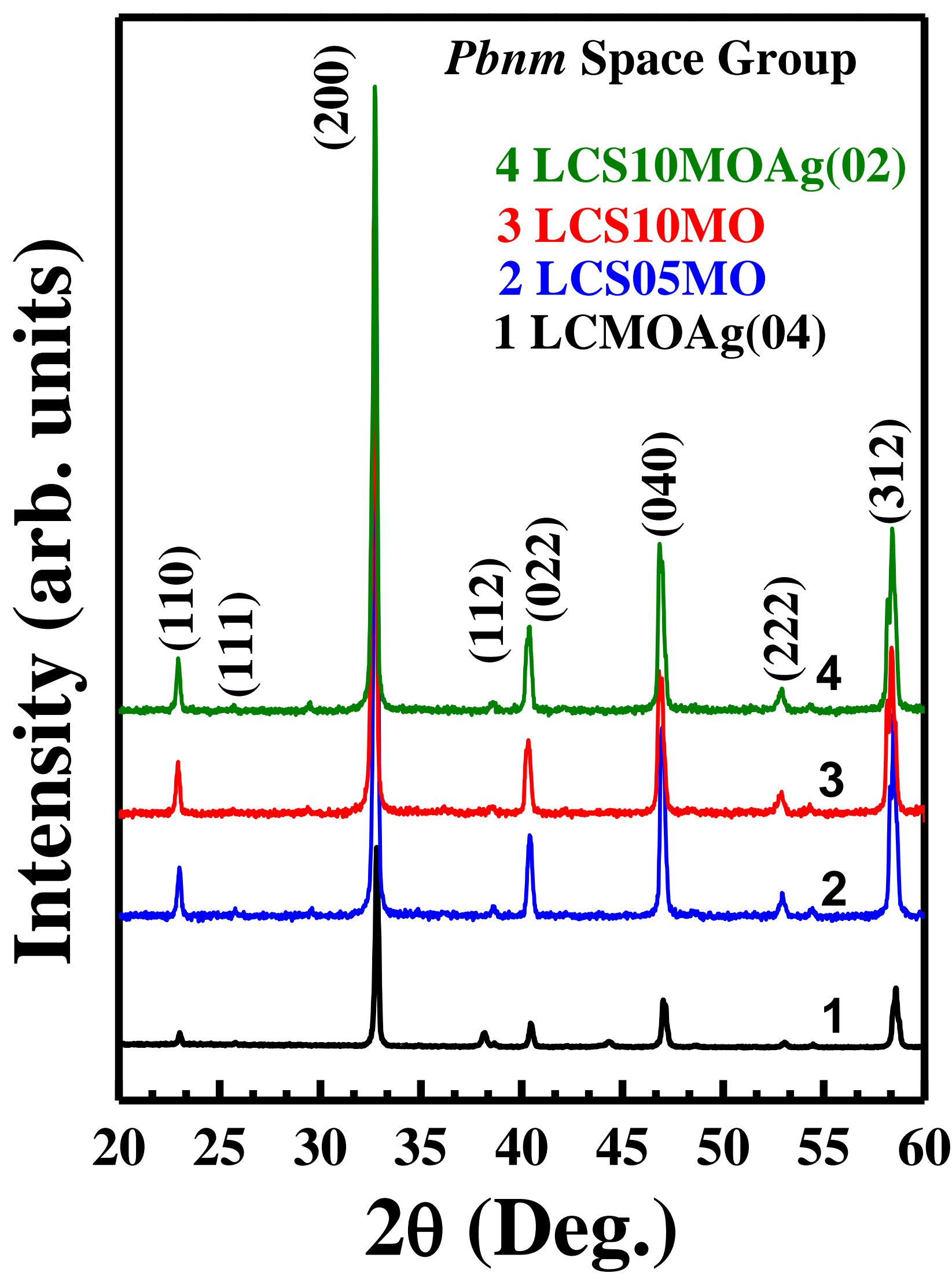

Figure 2

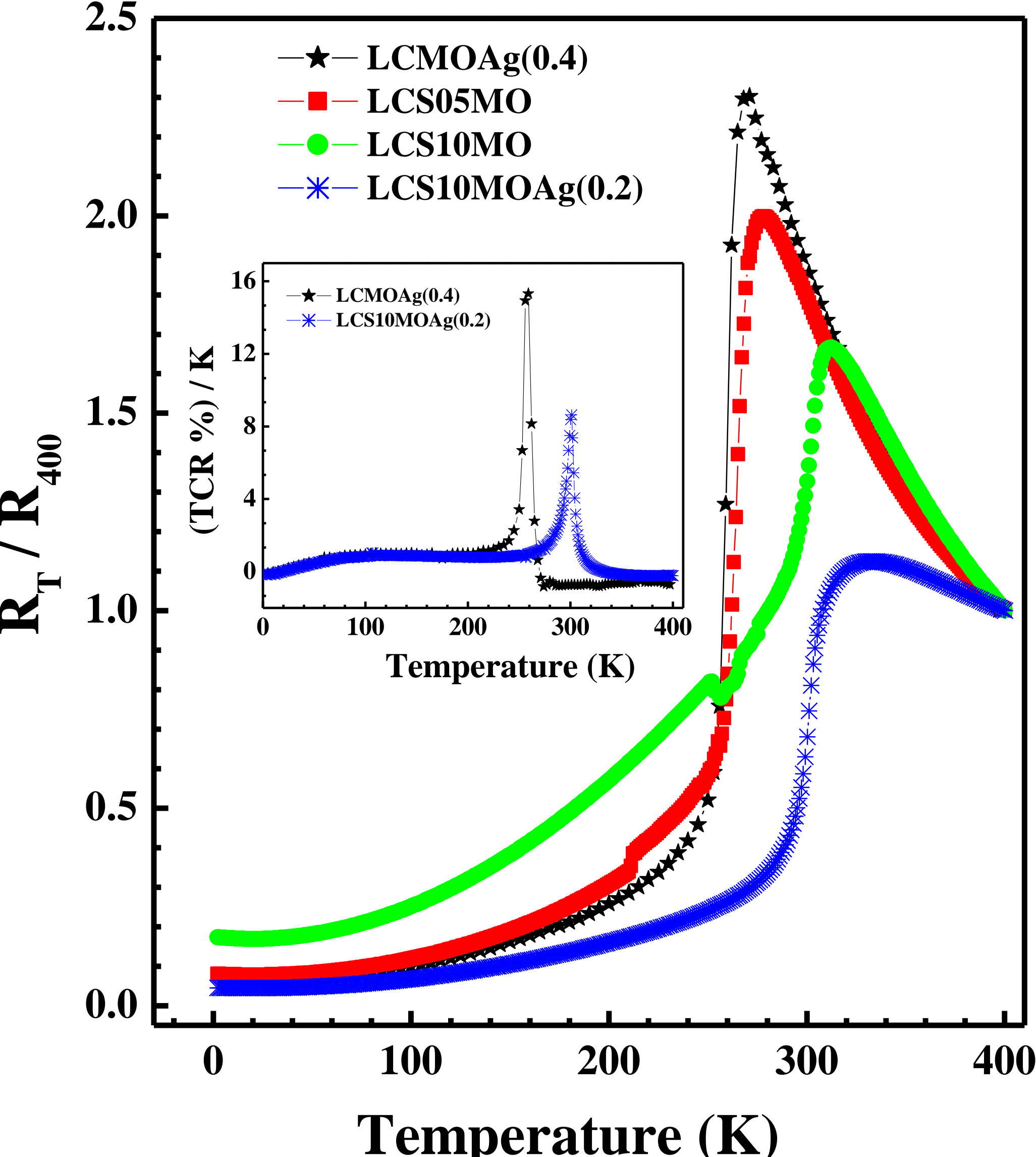

**Figure 3**
**(a)**

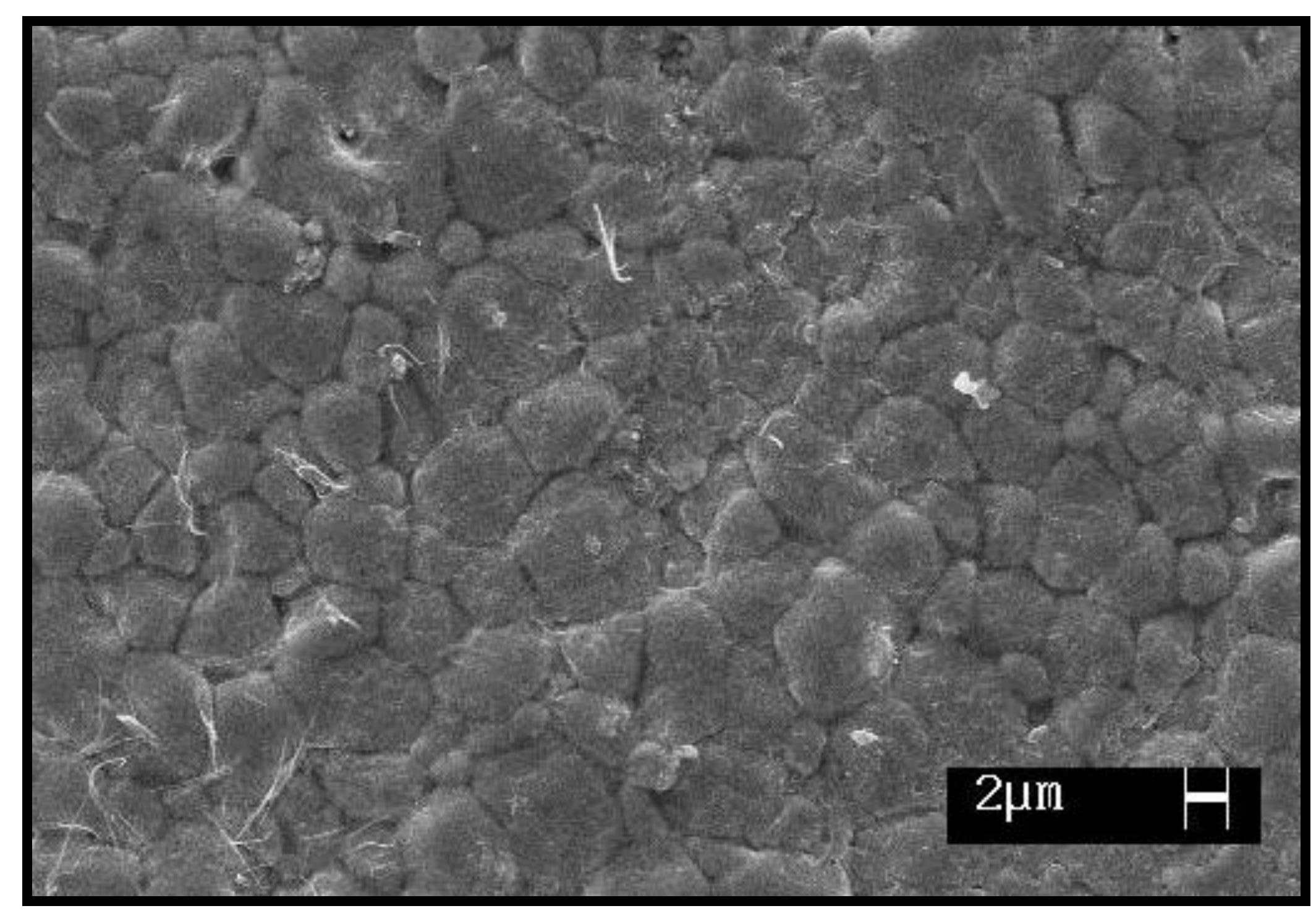

**(b)**

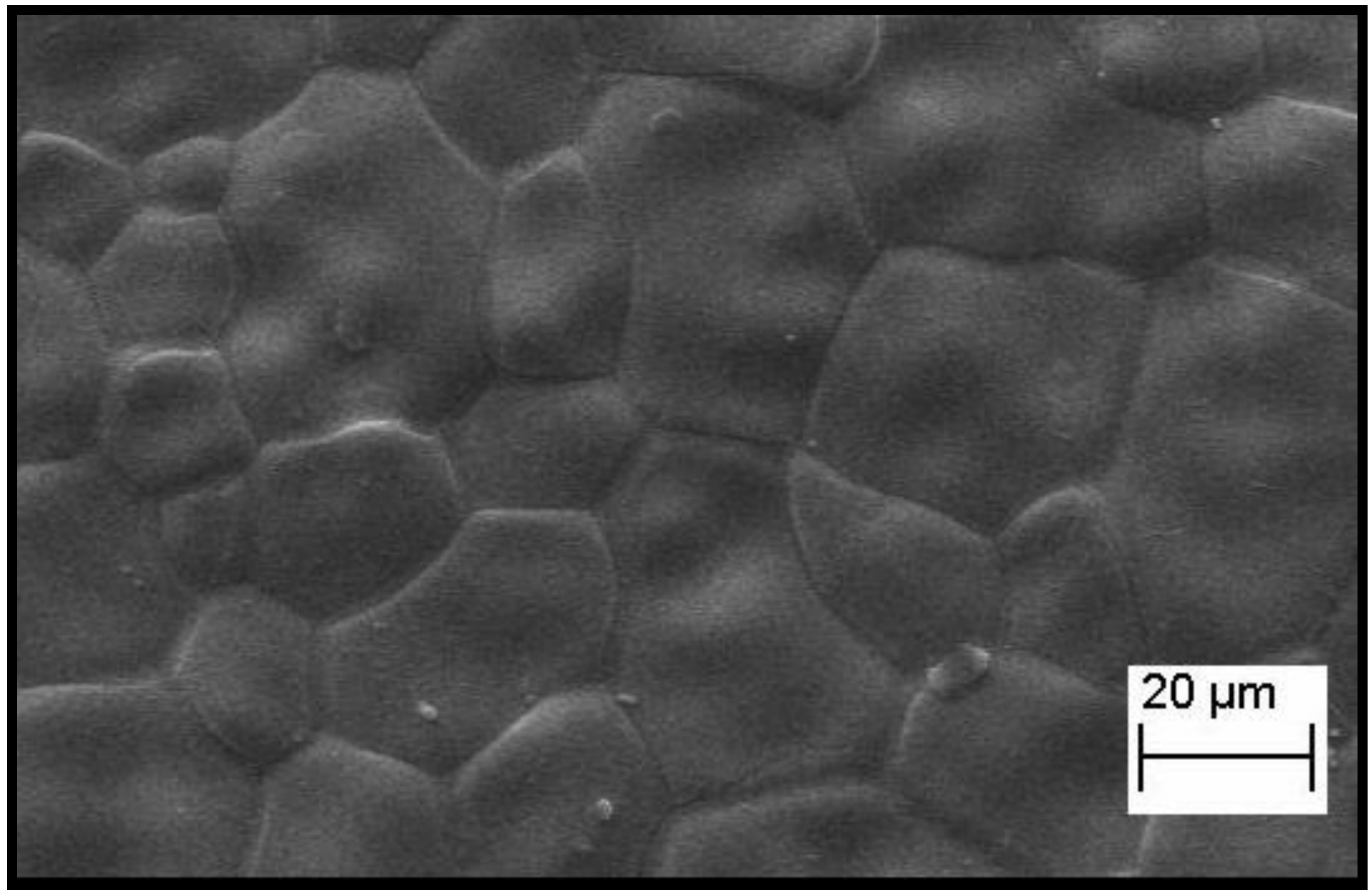

**Figure 4**

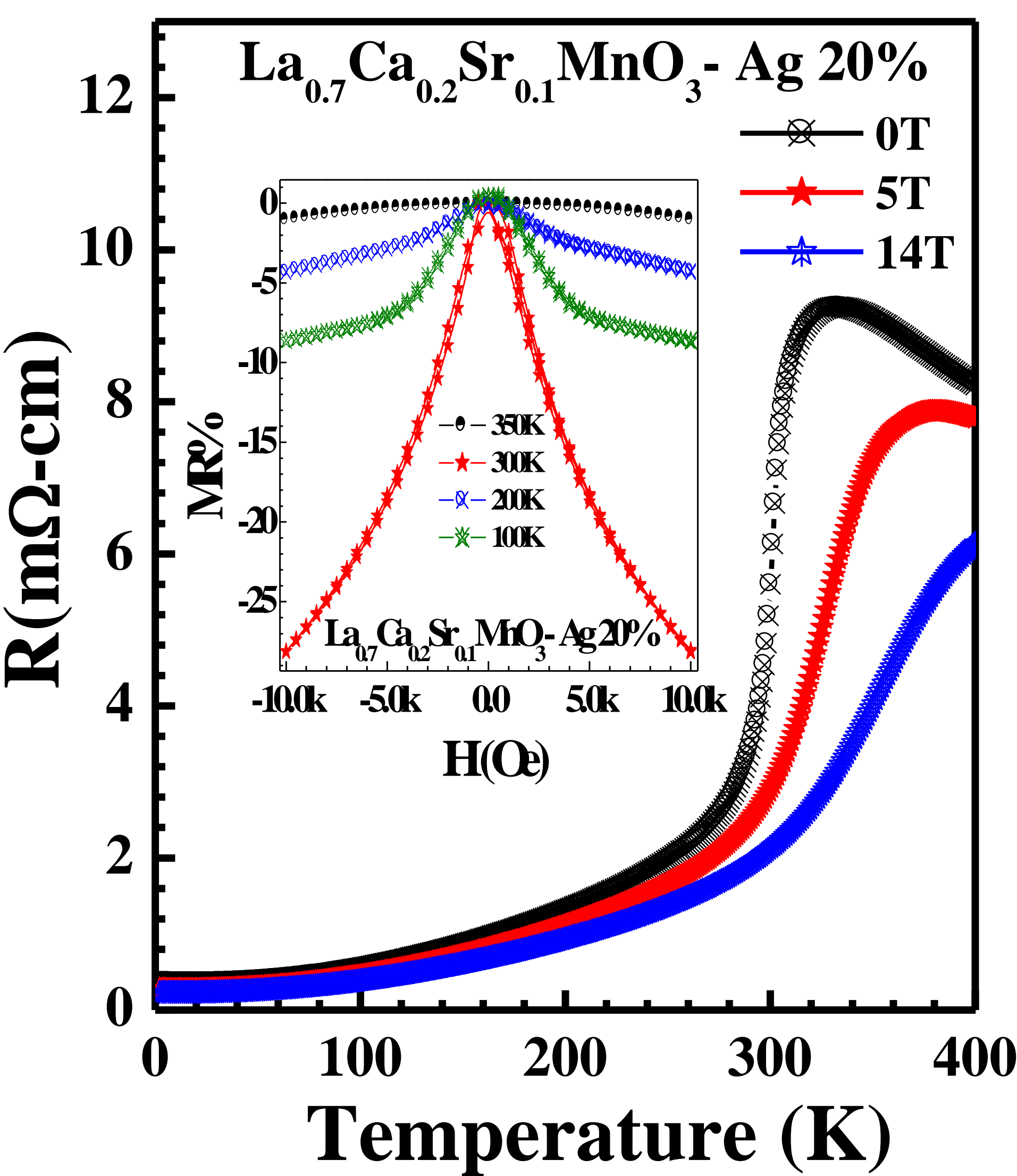

*******************************************